\documentclass[reprint,amsmath,amssymb,aps,superscriptaddress,showpacs,floatfix,byrevtex,longbibliography]{revtex4-1}

\usepackage{amsmath}
\usepackage{amssymb}
\usepackage{amstext}
\usepackage{amsopn}
\usepackage{amsfonts}
\usepackage{amsxtra}
\usepackage[english]{babel}
\usepackage{graphicx}
\usepackage{float}
\usepackage{bm}
\usepackage{multirow}
\usepackage{dcolumn}
\usepackage{color}
\usepackage{hyperref}

\begin{document}
%
% The title and the list of authors
%
\title{Perfect separation of intraband and interband excitations in PdCoO$_2$}
\author{C. C. Homes}
\email{homes@bnl.gov}
\affiliation{Condensed Matter Physics and Materials Science Division,
  Brookhaven National Laboratory, Upton, New York 11973, USA}%

\author{S. Khim}
\affiliation{Max Planck Institute for Chemical Physics of Solids, N\"{o}thnitzer Strasse 40,
  01187 Dresden, Germany}

\author{A. P. Mackenzie}
\email{andy.mackenzie@cpfs.mpg.de}
\affiliation{Max Planck Institute for Chemical Physics of Solids, N\"{o}thnitzer Strasse 40,
  01187 Dresden, Germany}
\affiliation{Scottish Universities Physics Alliance, School of Physics \& Astronomy,
  University of St.~Andrews, North Haugh, St.~Andrews KY16 9SS, United Kingdom}
\date{\today}

%
% The abstract goes here
%
\begin{abstract}
The temperature dependence of the optical properties of the delafossite PdCoO$_2$ has
been measured in the \emph{a-b} planes over a wide frequency range.  The optical conductivity
due to the free-carrier (intraband) response falls well below the interband transitions, allowing the
plasma frequency to be determined from the \emph{f}-sum rule.  Drude-Lorentz fits to the complex
optical conductivity yield estimates for the free-carrier plasma frequency and scattering rate.
The in-plane plasma frequency has also been calculated using density functional theory.
The experimentally-determined and calculated values for the plasma frequencies are
all in good agreement; however, at low temperature the optically-determined scattering rate
is much larger than the estimate for the transport scattering rate, indicating a strong
frequency-dependent renormalization of the optical scattering rate.  In addition to the
expected in-plane infrared-active modes, two very strong features are observed that are
attributed to the coupling of the in-plane carriers to the out-of-plane longitudinal optic modes.
\end{abstract}

%
%  PACS numbers
%  63.20.-e     Phonons in crystal lattices
%  72.15.Lh 	Relaxation times and mean free paths
%  75.30.-m 	Intrinsic properties of magnetically-ordered materials
% 78.20.Ci 	Optical constants
%  78.30.-j     Infrared and Raman spectra
%  78.30.Er 	Solid metals and alloys
%
\maketitle

%
% The main body of the text
%
% Introduction
%
\section{Introduction}
The delafossite PdCoO$_2$ is one of only a handful of transition metal oxides
whose in-plane resistivity at room temperature rivals that of silver or copper
($\simeq 2\,\mu\Omega\,{\rm cm}$), establishing a new benchmark for conducting
metal oxides \cite{shannon71}.   Perhaps even more remarkable is the large
resistivity ratio (${\rm RRR}\gtrsim 400$) and extremely low in-plane residual
resistivity at low temperature, $\rho_{ab}\simeq 8$~n$\Omega\,$cm \cite{tanaka96,hicks12},
which may place this material in the hydrodynamic limit \cite{moll16,zaanen16}.
Given the quasi two-dimensional (2D) behavior of this material \cite{tanaka96},
it is inevitable to compare it with the 2D cuprate materials, perhaps the most
studied of the conducting metal oxides.  The cuprates are typically described as bad
metals \cite{emery95} in which the resistivity often shows a peculiar non-saturating
linear temperature dependence that may violate the Mott-Ioffe-Regel limit at high
temperature\cite{hussey04,MIR}; the optical conductivity reveals an unusual free-carrier
response where the scattering rate is strongly renormalized with frequency, resulting
in an incoherent response that merges with other bound excitations \cite{basov05}.
Surprisingly, in both of these materials the free carriers originate from a single
band at the Fermi level.
The common structural motif in the cuprates is the square copper-oxygen
plaquettes where the conducting states originate; however, PdCoO$_2$ is
different in that it crystalizes in the trigonal $R\bar{3}m$ (166) space
group, consisting of Pd triangular layers and CoO$_2$ triangular slabs
\cite{shannon71}, shown in the inset of Fig.~\ref{fig:reflec}.
There is theoretical \cite{seshadri98,eyert08,kim09,ong10} as well as
experimental \cite{noh09} evidence that the density of states at the
Fermi level is dominated by Pd rather than Co, indicating that the
conduction originates in the Pd layers.
The exceptionally long in-plane mean free paths of $\simeq 20\,\mu$m at
low temperature implies that the Pd layers are almost completely free of
any disorder, since the mean free path corresponds to $\simeq 10^{5}$ lattice
spacings \cite{mackenzie17}, a situation that is difficult to justify
given that the crystals are grown using flux-based techniques.

%
% This work
%
%In this Letter the complex optical properties of PdCoO$_2$ have been determined for
%the first time over a wide spectral range for a variety of temperatures.  The
%real part of the optical conductivity is particularly useful as it yields information
%about the free-carrier response and interband transitions; in the zero-frequency limit,
%the dc conductivity is recovered, allowing comparisons to be made with transport data.
%Furthermore, the infrared-active transverse-optic modes at the center of the
%Brillouin zone may be observed as resonances superimposed upon an electronic
%background (or antiresonances if strong electron-phonon coupling is present).

In this work the complex optical properties of PdCoO$_2$ have been determined for light
polarized in the \emph{a-b} planes over a wide frequency range at a variety of temperatures.
The real part of the optical conductivity reveals that the free-carrier response is completely
isolated from the interband transitions, allowing the plasma frequency to be determined
from the \emph{f}-sum rule.  The free-carrier response has also been fit using the Drude-Lorentz
model, returning values for the plasma frequency and scattering rate.  In addition,
the in-plane and out-of-plane plasma frequencies and interband optical conductivities
have been calculated using density functional theory.  The calculated and
experimentally-determined plasma frequencies are all in good agreement.  However,
at low temperature the experimentally-determined optical scattering rate is much
larger than the estimated transport scattering rate; this disagreement may only be
resolved if the optical scattering rate is assumed to vary quadratically with frequency
(Fermi liquid), as opposed to the linear dependence observed in the cuprates
(marginal Fermi liquid).  Finally, in addition to the expected in-plane infrared-active
modes, two very strong features are attributed to the coupling of the in-plane carriers
with the out-of-plane longitudinal optic (LO) modes \cite{reedyk92}, indicating the
presence of electron-phonon coupling.

%
% Experiment and results
%
% Crystal growth here.
%
\section{Experiment}
Single crystals of PdCoO$_2$ were grown in an evacuated quartz ampoule with
a mixture of PdCl$_2$ and CoO as described in Refs.~\onlinecite{shannon71,takatsu10},
yielding thin platelets of typical dimensions $1\,{\rm mm}\times 1\,{\rm mm}
\times 100\,\mu{\rm m}$.
The reflectance of PdCoO$_2$ was measured at a near-normal angle of incidence for
light polarized in the \emph{a-b} planes at a variety of temperatures over a wide
frequency range ($\simeq 5$~meV to 5~eV) using an overfilling and \emph{in situ}
evaporation technique \cite{homes93}.
While the reflectance contains a great deal of information, it is a combination of the
real and imaginary parts of the refractive index, and as such it is not an intuitive
quantity.  The complex optical properties have been calculated from a Kramers-Kronig
analysis of the reflectance \cite{dressel-book,wooten}, which requires extrapolations
for $\omega\rightarrow 0,\infty$.  Below the lowest-measured frequency point, a
Hagen-Rubens form is employed, $R(\omega)\propto 1-a\sqrt{\omega}$, where $a$ is
chosen to match the data.  Above the highest-measured frequency, the reflectance is
assumed to have the power-law dependence $R(\omega)\propto 1/\omega$ up to $1.5\times
10^5$~cm$^{-1}$, above which a free-electron $1/\omega^4$ behavior is assumed.

\section{Results and Discussion}
\begin{figure}[t]
\includegraphics[width=3.1in]{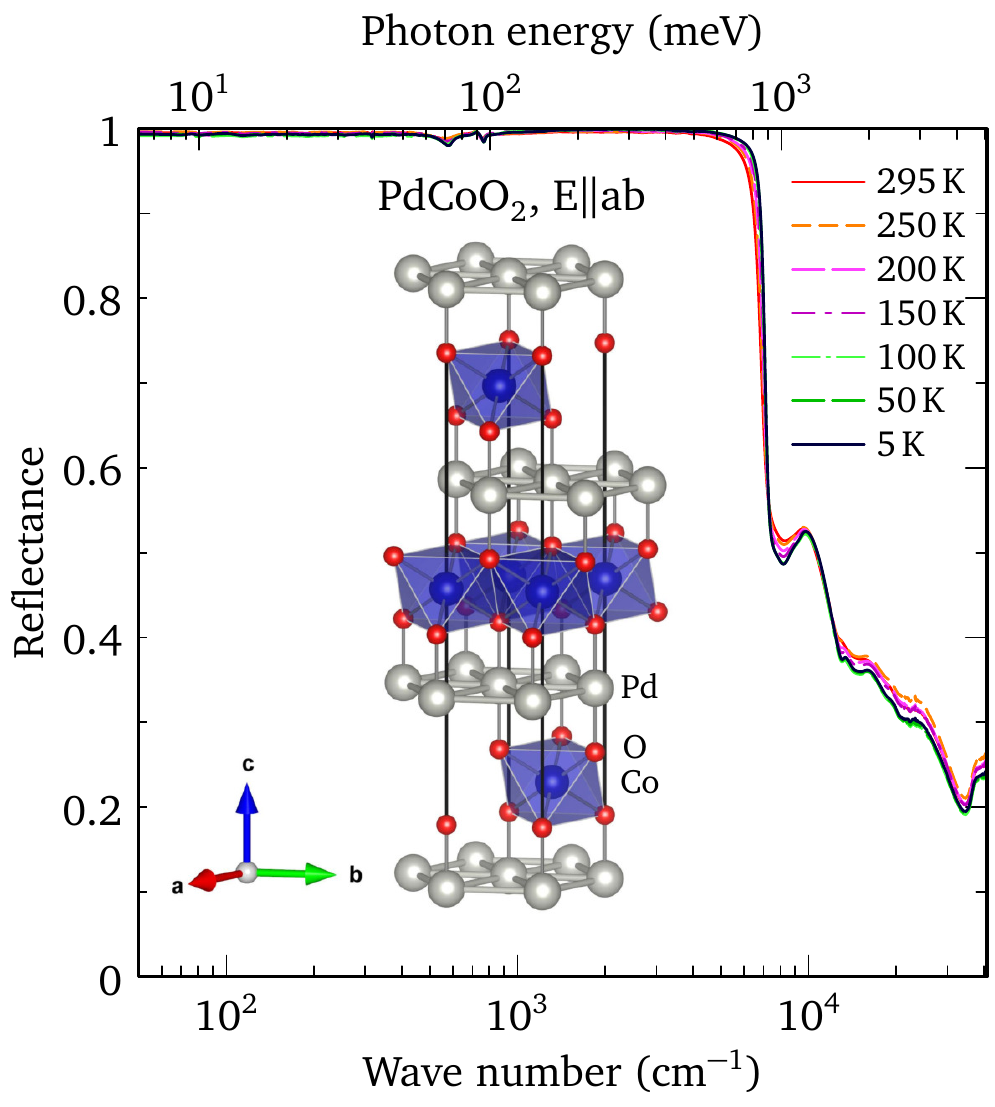}
\caption{The temperature dependence of the reflectance of PdCoO$_2$ for light
polarized in the \emph{a-b} planes showing its extremely high value over the
far and mid-infrared regions until a sharp plasma edge is encountered at
$\simeq 6000$~cm$^{-1}$.
Inset: The unit cell of PdCoO$_2$ depicting the triangular coordination of the
Pd atoms and the CoO$_2$ slabs within the \emph{a-b} planes \cite{vesta}.}
%
%\vspace*{-0.0cm}%
\label{fig:reflec}
\end{figure}

The temperature dependence of the reflectance is shown in over a wide frequency
range in Fig.~\ref{fig:reflec}; a remarkable feature of the reflectance is its
extremely high value ($\gtrsim 0.99$) over the far- and mid-infrared regions,
with a sharp plasma edge at $\simeq 6000$~cm$^{-1}$ ($\simeq 0.7$~eV).
We note that this measurement is particularly challenging because at room temperature
the in-plane reflectance of PdCoO$_2$ in this region is already higher than that of
gold or silver \cite{palik85}, two elements that are used as optical references.
Despite the dramatic decrease in the resistivity at low temperature \cite{hicks12},
the only noticeable change in the reflectance is a slight sharpening of the plasma edge.
Interestingly, there is also structure in the $500 - 800$~cm$^{-1}$ region, the energy
range associated with lattice vibrations.
%

%
% Figure 1: Reflectance
%

\subsection{Complex conductivity}
\begin{figure}[t]
\includegraphics[width=3.1in]{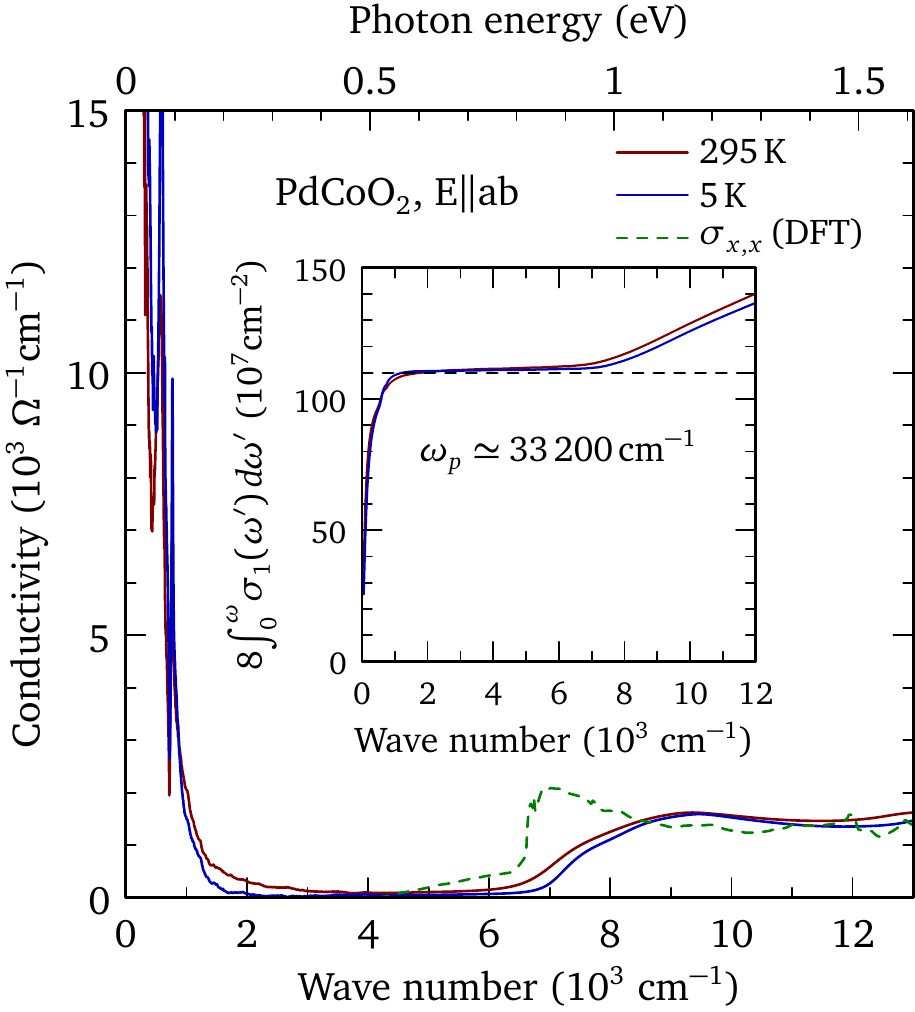}
\caption{The real part of the optical conductivity of PdCoO$_2$ for light polarized
in the \emph{a-b} planes at 295 and 5~K showing the complete separation of the
low-frequency intraband excitations below $\simeq 1500$~cm$^{-1}$, and the interband
transitions, which have an onset at $\simeq 7000$~cm$^{-1}$.  The calculated optical
conductivity due to the interband transitions is denoted by $\sigma_{x,x}$.
Inset: The spectral weight associated with the conductivity sum rule at 295 and 5~K.}
%
%\vspace*{-0.0cm}%
\label{fig:sigma}
\end{figure}

The real part of the optical conductivity is shown over a wide spectral region
at 295 and $\simeq 5$~K in Fig.~\ref{fig:sigma}.   Interestingly, at low
temperature the low-frequency conductivity associated with the free-carrier
response is limited to below $\simeq 1500$~cm$^{-1}$; as a consequence, there
is no overlap with the interband transitions which are observed above $\simeq 7000$~cm$^{-1}$.
%
% f-sum rule
%
The \emph{f}-sum rule allows that in the absence of other excitations,
$\int_0^{\omega} \sigma_1(\omega^\prime)d\omega^\prime = \omega_p^2/8$, where
$\omega_p^2 = 4\pi ne^2/m^\ast$ is the square of plasma frequency with carrier
concentration $n$ and effective mass $m^\ast$, and the cut-off frequency $\omega$ is
chosen so that $\omega_p$ converges smoothly (here $\sigma_1$ has the units of cm$^{-1}$).
The inset in Fig.~\ref{fig:sigma} shows the result of this conductivity sum rule
up to $\sim 1.5$~eV; the integral has converged by about $\omega \simeq 1000$~cm$^{-1}$,
yielding a value of $\omega_p \simeq 33\,200\pm 600$~cm$^{-1}$ at both 295 and 5~K;
this is close to the value of $\omega_p\simeq 38\,200$~cm$^{-1}$ determined from a
de Haas--van Alphen study \cite{hicks12}.

%
% Figure 2: real part of the optical conductivity
%

The optical conductivity may also be fit to the Drude-Lorentz model with the complex
dielectric function $\tilde\epsilon = \epsilon_1 + i\epsilon_2$,
\begin{equation}
  \tilde\epsilon(\omega) = \epsilon_\infty - {{\omega_{p}^2}\over{\omega^2+i\omega/\tau_{op}}}
    + \sum_j {{\Omega_j^2}\over{\omega_j^2 - \omega^2 - i\omega\gamma_j}},
  \label{eq:dl}
\end{equation}
where $\epsilon_\infty$ is the real part of the dielectric function at high frequency,
$\omega_{p}$ is previously defined, and $1/\tau_{op}$ is the scattering rate for the delocalized
(Drude) carriers; typically, $1/\tau_{op}$ is nearly identical to the scattering rate
determined from transport measurements, $1/\tau_{tr}$.  In the summation, $\omega_j$, $\gamma_j$
and $\Omega_j$ are the position, width, and strength of the $j$th transverse optic (TO) mode
or a bound excitation, respectively.  The complex conductivity is $\tilde\sigma = \sigma_1+i\sigma_2
= 2\pi i\omega[\epsilon_\infty - \tilde\epsilon(\omega)]/Z_0$ (in units of $\Omega^{-1}$cm$^{-1}$),
where $Z_0=377$~$\Omega$ is the impedance of free space.
%
% Figure 3: real part of the optical conductivity
%
\begin{figure}[t]
\includegraphics[width=3.15in]{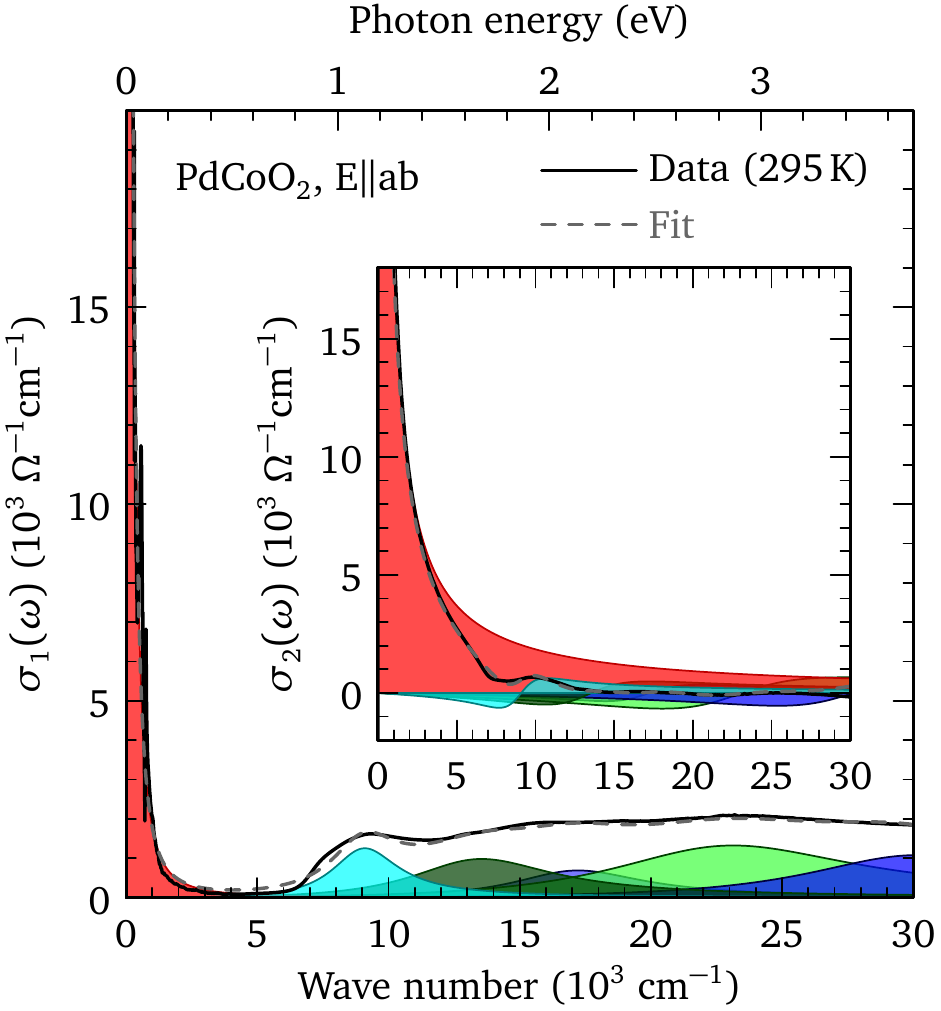}
\caption{The real part of the optical conductivity of PdCoO$_2$ at 295~K for
light polarized in the \emph{a-b} planes (solid line), compared to the
fit to the Drude-Lorentz model (dashed line), which is decomposed into the individual
Drude and Lorentz oscillator components; the overall quality of the fit is quite good
(see Table~\ref{tab:complex} for parameter values).
Inset: The imaginary part of the optical conductivity (solid line), compared to the fitted
value (dashed line).  Compared to the real part, the free-carrier response of the imaginary
part is considerably broader.}
%
%\vspace*{-0.0cm}%
\label{fig:fits}
\end{figure}
The results of the simultaneous fit to the real and imaginary parts of the optical conductivity
of PdCoO$_2$ at 295~K for light polarized in the \emph{a-b} planes with a single Drude component
and five Lorentz oscillators using a non-linear least-squares technique is shown in Fig.~\ref{fig:fits};
the values of the fitted parameters are listed in Table~\ref{tab:complex}.
The fit to the real part of the optical conductivity returns a Drude component with
$\omega_p\simeq 33\,300$~cm$^{-1}$ and $1/\tau_{op}\simeq 97$~cm$^{-1}$.  While
the result for the plasma frequency is in excellent agreement with the value
determined from the \emph{f}-sum rule, the value for $1/\tau_{op}$ is more than an
order of magnitude smaller than the scattering rate suggested from photoemission
experiments \cite{noh09}.
The corresponding fit to the imaginary part of the optical conductivity in the inset
of Fig.~\ref{fig:fits} indicates that the free-carrier response extends well into the
mid-infrared region and allows the high-frequency part of the dielectric function to be
determined, $\epsilon_\infty\simeq 3.4$.
A minimal number of oscillators has been used to describe the relatively flat optical
conductivity at high frequency; however, it should be noted that the placement of the
high-frequency mode ($\omega_5$) is somewhat arbitrary.

%
% Scattering rates, etc.
%
\subsection{Scattering rates}
Using the Drude expression for the dc conductivity $\sigma_{0} \equiv
\sigma_1(\omega\rightarrow 0) =  2\pi\omega_{p}^2\tau_{op}/Z_0$ with the
values for the fitted parameters at 295~K yields $\sigma_{0} \simeq 1.85
\times 10^5$~$\Omega^{-1}$cm$^{-1}$, or in terms of the resistivity,
$\rho_{0}\simeq 5.4$~$\mu\Omega\,$cm, which is close to the transport
value, $\rho_{dc} \simeq 2.6$~$\mu\Omega\,$cm \cite{hicks12}.
Given that $\rho_{ab}(300\,\rm{K})/\rho_{ab}(2\,{\rm K})\simeq 400$ \cite{takatsu07},
and $\rho_{dc} \propto 1/\tau_{tr}$, then at low temperature $1/\tau_{tr} \lesssim 0.3$~cm$^{-1}$;
however, fits to both the optical conductivity and the plasma edge in the
reflectance at $\simeq 5$~K yield the significantly larger values of $1/\tau_{op}
\simeq 80\pm 10$~cm$^{-1}$.  While a value for $1/\tau_{op} \lesssim 1$~cm$^{-1}$ might
reasonably be thought to result in a plasma edge in the reflectance that resembles a step
function, the proximity to nearby interband transitions has the effect of
significantly broadening this feature.

\begin{figure}[t]
\includegraphics[width=3.1in]{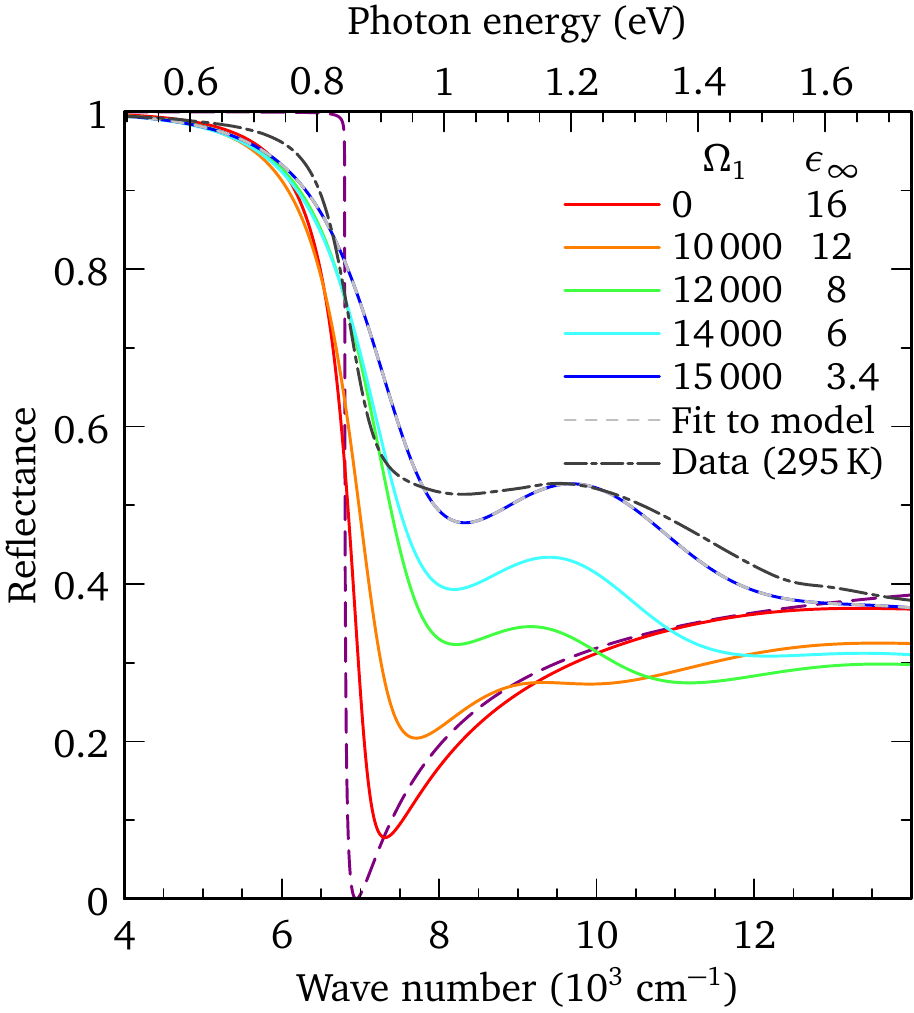} \
\caption{The experimentally-determined reflectance of PdCoO$_2$ for light polarized in the
\emph{a-b} planes at 295~K (dash-dot line), compared to the calculated reflectance for free
carriers with $\omega_{p}=33\,500$~cm$^{-1}$, $1/\tau_{op}=1$~cm$^{-1}$, and $\epsilon_\infty=25$
in the absence of any other excitations (dashed line), and in the presence of an interband
transition with $\omega_1=9100$~cm$^{-1}$, $\gamma_1=3100$~cm$^{-1}$, and varying strengths
$\Omega_1=0\rightarrow 15\,000$~cm$^{-1}$ ($\epsilon_\infty =16 \rightarrow 3.4$),
along with four other high-frequency modes (Table~\ref{tab:complex}).  The Drude-Lorentz fit
to the upper reflectance curve reproduces the model perfectly (Table~\ref{tab:fits}).
}
\label{fig:plasma}
\end{figure}

To demonstrate this effect, the reflectance $R=\tilde{r}\tilde{r}^\ast$ has been calculated
at a normal angle of incidence; $\tilde{r}=(\tilde{n}-1)/(\tilde{n}+1)$ is the Fresnel
reflectance, which is related to the dielectric function through the complex refractive
index, $\tilde{\epsilon} = \tilde{n}^2=(n+ik)^2$.  The Drude reflectance is initially calculated
in the absence of any interband excitations for $\omega_{p}=33\,500$~cm$^{-1}$ with a scattering
rate of $1/\tau_{op}=1$~cm$^{-1}$, and $\epsilon_\infty=25$ (the value of $\epsilon_\infty$ is
chosen to place the renormalized plasma frequency, $\omega_p/\sqrt{\epsilon_\infty}$,
close to the experimentally-observed position); the resulting plasma edge is extremely sharp with
a sharp drop at just over 0.8~eV that resembles a step function, shown by the dashed line in
Fig.~\ref{fig:plasma}.

The five interband excitations in Table~\ref{tab:complex} have been added to the reflectance; however,
the low-frequency oscillator is described by $\omega_1=9100$~cm$^{-1}$, with width $\gamma_1=3100$~cm$^{-1}$,
and a gradually increasing value of the oscillator strength of $\Omega_1 = 0 \rightarrow 15\,000$~cm$^{-1}$
(the values of $\epsilon_\infty$ have been adjusted by hand to keep the value of the reflectance at high
frequency roughly constant); the addition of the interband terms, the low-frequency oscillator in
particular, has the effect of broadening the Drude plasma edge considerably, as well as shifting the
minima to higher frequency.  The broadened nature of the plasma edge in the reflectance might make it
difficult to determine the intrinsic value of the free-carrier scattering rate.

%
% Table I
%
\begin{table}[t]
\caption{The Drude-Lorentz model parameters fitted to the real and imaginary parts
of the optical conductivity of PdCoO$_2$ at 295~K for light polarized in the \emph{a-b}
planes. All values are in units of cm$^{-1}$, unless otherwise indicated.$^a$}
  \begin{ruledtabular}
  \begin{tabular}{cc ccc}
   Component  &   $j$  & $\omega_j$  & $1/\tau_{op}$, $\gamma_j$ & $\omega_p$, $\Omega_j$ \\
  \cline{1-5}
  Drude   &   &   $-$  &    97 &  33317 \\
  Lorentz & 1 &  9099  &  3067 &  15200 \\
  Lorentz & 2 & 13555  &  6620 &  19760 \\
  Lorentz & 3 & 17223  &  6085 &  15880 \\
  Lorentz & 4 & 23135  & 11585 &  30360 \\
  Lorentz & 5 & 30200  & 10557 &  26190 \\
  \end{tabular}
  \end{ruledtabular}
  \footnotetext[1] {$\epsilon_\infty = 3.4$.}
  \label{tab:complex}
\end{table}

To test this possibility, we have fit the upper curve in Fig.~\ref{fig:plasma}
that most closely resembles the experimental reflectance at 295~K (dash-dot line) using the model
values in Table~\ref{tab:fits} (the remaining high-frequency oscillators are taken from
Table~\ref{tab:complex}), employing a spectral resolution of 8~cm$^{-1}$ (less than or
equal to the experimental resolution in the mid- and near-infrared regions).  The free-carrier
component and the low-frequency oscillator are fit to the reflectance using a non-linear
least-squares method, while the four high-frequency modes are kept fixed; the fitted results are
identical to the model values, which are summarized in Table~\ref{tab:fits}.
This indicates that despite the broadening of the plasma edge in the reflectance due to nearby
interband transitions, as well as an instrumental resolution that is lower than the intrinsic
width of this feature, the values for $1/\tau_{op} \lesssim 1$~cm$^{-1}$ may still be accurately
determined from fits to the reflectance (or complex conductivity).

%
% Figure 4: calculated reflectance
%

Thus, we conclude that there is a profound disagreement between the optical and
transport scattering rates at low temperature, $1/\tau_{op} \gg 1/\tau_{tr}$.
This discrepancy may arise if within a single band the optical scattering rate
is strongly renormalized with frequency, as described by the generalized Drude
model  \cite{dordevic06,homes15}
\begin{equation}
  1/\tau_{op}(\omega) = \frac{2\pi\omega_p^2}{Z_0}\, \Re\left[ \frac{1}{\tilde\sigma(\omega)}
\right].
\end{equation}
However, this approach is complicated by the fact that $\sigma_1(\omega)\propto
1/\left[ 1-R(\omega)\right]$; because the reflectance is close to unity, even a
small uncertainty can result in large changes to $\sigma_1$, and subsequently the
scattering rate, making the experimentally-determined values unreliable.

While the in-plane transport at very low temperatures is rather unusual and perhaps
governed by phonon-drag effects \cite{hicks12}, in general this material may be regarded
as a Fermi liquid \cite{tanaka96}, where the scattering rate is quadratic in both
temperature and frequency,
$1/\tau(\omega, T) = 1/\tau_0 + a(\hbar\omega)^2 + b(k_{\rm B}T)^2$,
where $b/a=\pi^2$ \cite{dordevic06}.  In the $\omega\rightarrow 0$ (dc) limit
$1/\tau(T) = 1/\tau_0 + b(k_{\rm B}T)^2$, with a residual scattering rate
$1/\tau_0 \simeq 0.14$~cm$^{-1}$ \cite{hicks12}.
Using the Drude scattering rate $1/\tau_{op} \simeq 100$~cm$^{-1}$ at 295~K we can estimate
$b\simeq 2.383\times 10^{-3}$~cm \footnote{A useful conversion is 1~eV = 8065.5~cm$^{-1}$}.
In the frequency domain at low temperature, $1/\tau_{op}$ is then the average of $1/\tau(\omega)$
over the interval $0\rightarrow \omega$,
\begin{equation}
  1/\tau_{op} = \frac{1}{\omega} \int_0^{\omega} 1/\tau(\omega^\prime)\,d\omega^\prime
  \simeq \frac{b}{3\pi^2}\omega^2 .
\end{equation}
From Fig.~\ref{fig:sigma}, a reasonable estimate for $\omega$ would be the point at
which most of the spectral weight from the free carriers is captured, $\omega \simeq
1000$~cm$^{-1}$, resulting in $1/\tau_{op} \simeq 80$~cm$^{-1}$, which is in
excellent agreement with the Drude estimates for the optical scattering rate at
low temperature.

%
% Table II
%
\begin{table}[t]
\caption{Initial parameters for the Drude-Lorentz model, as well as the seed and final
values for the fit to the model reflectance.$^a$  The high-frequency oscillators used in
the model, $\omega_2$ through $\omega_5$ (Table~\ref{tab:complex}) are kept fixed.
All units are in cm$^{-1}$.}
  \begin{ruledtabular}
  \begin{tabular}{cccc}
   Parameter  & Model value  & Seed value & Fitted value \\
  \cline{1-4}
   $\omega_{p}$     & $33\,500$  & $30\,000$  & $33\,500$  \\
   $1/\tau_{op}$    & 1.0        & 100        & 1.00  \\
%
% $\epsilon_\infty$ &    3.4     &  5         & 3.4 \\
%
   $\omega_1$        & 9100       & 9000       & 9100  \\
   $\gamma_1$        & 3100       & 2000       & 3100  \\
   $\Omega_1$        & $15\,000$  & $10\,000$  & $15\,000$  \\
  \end{tabular}
  \end{ruledtabular}
  \footnotetext[1] {The model value is $\epsilon_\infty = 3.4$; the seed value is 5,
  the fitted value is 3.4.}
  \label{tab:fits}
\end{table}

%
% DFT calculations
%
\subsection{Electronic structure}
Several first principle calculations have been undertaken to study the
electronic \cite{seshadri98,eyert08,kim09,ong10} and vibrational \cite{kumar13,cheng17}
properties of PdCoO$_2$; however, we are unaware of any that have dealt with the optical properties.
Accordingly, the electronic properties have been calculated using density functional
theory (DFT) with the generalized gradient approximation (GGA) using the full-potential
linearized augmented plane-wave (FP-LAPW) method \cite{singh} with local-orbital
extensions \cite{singh91} in the WIEN2k  implementation \cite{wien2k}.  The total energy
and residual forces have been minimized with respect to the unit cell parameters and the
fractional coordinates, respectively (details are provided in the Supplementary Material
\cite{suplmt}).
%
% An examination of different Monkhorst-Pack {\em k}-point meshes indicated that a $5\times{5}\times{5}$
% {\em k}-point mesh with $R_{mt}k_{max}=8$ was sufficient for good energy convergence.
% Beginning with the experimental unit cell \cite{shannon71}, the total energy was minimized
% by adjusting $a$ and $c$ axes; the atomic fractional coordinate for the oxygen atom was then
% relaxed with respect to the total force, typically resulting in residual forces of less than
% 0.1~mRy/a.u.~per atom.  This procedure was repeated until no further improvement was obtained.
% For both of these procedures spin-orbit coupling is ignored.  The final values for the unit cell
% are listed in Table~\ref{tab:dft}
%
The real part of the optical conductivity including the effects of spin orbit coupling
has been calculated from the imaginary part of the dielectric function,
$\sigma_{x,x} = 2\pi\omega\,\Im\,\epsilon_{x,x}/Z_0$ \cite{draxl06}, using a fine $k$-point
mesh ($10\,000$~$k$ points).  The calculated conductivity due to interband transitions
along the \emph{a} axis ($\sigma_{x,x}$) shown in Fig.~\ref{fig:sigma} is in excellent
agreement with the experimental results.  The intraband plasma frequencies have also been
calculated for the \emph{a} and \emph{c} axes with values of $\omega_{p,a}\simeq 31\,500$~cm$^{-1}$
and $\omega_{p,c}\simeq 3660$~cm$^{-1}$, respectively, indicating a large anisotropy in the effective
mass $\omega_{p,a}^2/\omega_{p,c}^2=m_c^\ast/m_a^\ast \simeq 74$; this is consistent with the
quasi-2D nature of this material.  The value for $\omega_{p,a}$ is in good agreement with
the experimentally-determined average in-plane value of $\omega_p\simeq 33\,300$~cm$^{-1}$.

%
% Figure 5: detailed low-frequency conductivity
%
\begin{figure}[t]
\includegraphics[width=3.1in]{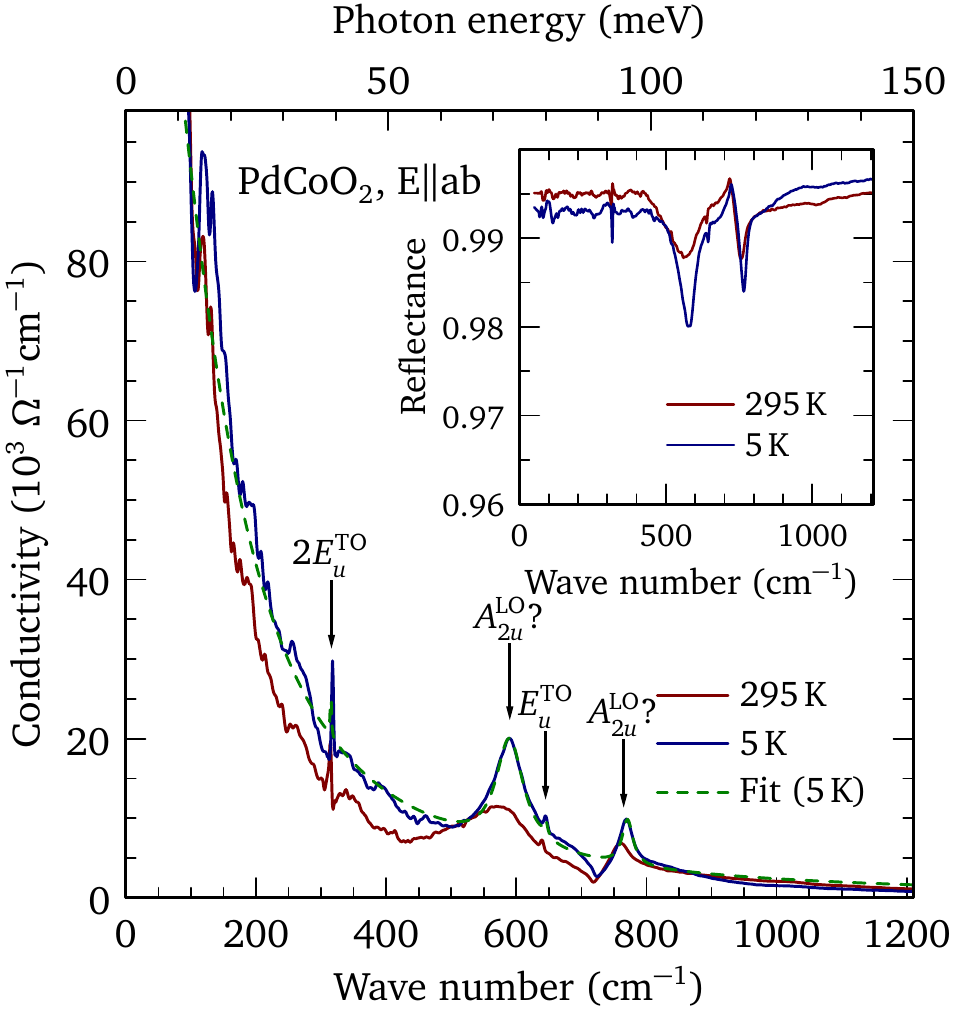}
\caption{The real part of the optical conductivity of PdCoO$_2$ for light polarized
in the \emph{a-b} planes at 295 and 5~K in the low-frequency region showing several
strong features superimposed on the free-carrier response.  The dashed line is the
Drude-Lorentz fit to the data at 5~K.
Inset: The reflectance at 295 and 5~K over the same frequency interval.}
%
%\vspace*{-0.0cm}%
\label{fig:modes}
\end{figure}
%

%
% The resistivity anisotropy may be expressed in terms of the optically-determined plasma
% frequencies and scattering rates,
%
%\begin{equation}
%  \frac{\rho_c}{\rho_{ab}} = \frac{\omega_{p,a}^2\tau_{op,a}}{\omega_{p,c}^2\tau_{op,c}}.
%\end{equation}
%
%At room temperature, $\rho_c/\rho_{ab} \simeq 410$ \cite{mackenzie17}; using the
%result $\omega_{p,a}^2/\omega_{p,c}^2=m_c^\ast/m_a^\ast \simeq 74$, we can infer that
%$\tau_{op,a}\simeq 5.6\,\tau_{op,c}$, increasing to $\tau_{op,a}\simeq 15\,\tau_{op,c}$
%at low temperature.

%
% Vibrational features
%
\subsection{Vibrational properties}
The low-frequency optical conductivity in Fig.~\ref{fig:sigma} has some structure
superimposed on the free-carrier response.  The optical conductivity at
295 and 5~K is shown in Fig.~\ref{fig:modes} below $\sim 0.15$~eV; several very strong
features are observed, which are also present in the reflectance (shown in the inset).
The irreducible vibrational representation for PdCoO$_2$ for the $R\bar{3}m$
space group is $\Gamma_{vib}=A_{1g}+E_g+2A_{2u}+2E_u$; the $A_{1g}$ and $E_g$ modes are
Raman active, while the $A_{2u}$ and $E_u$ modes are infrared-active along the \emph{c}
and \emph{a} axes, respectively \cite{takatsu07}.  While only two infrared-active
vibrations are expected for light polarized in the \emph{a-b} planes, it is clear
from Fig.~\ref{fig:modes} that there are at least four modes present.  The features
in the optical conductivity have been fit to Lorentzian oscillators [Eq.~(\ref{eq:dl})],
and the results shown in Table~\ref{tab:modes}.

There are several existing first-principles calculations of the lattice modes in this
material \cite{kumar13,cheng17}, which we have reproduced using the frozen-phonon
(direct) method to determine the atomic character of the zone-center TO vibrations
\cite{phonon} (details are provided in the Supplementary Material); the results are
summarized in Table~\ref{tab:modes}.
The low-frequency $E_u$ mode involves mainly the Pd and Co atoms and is
calculated to be at $\simeq{154}$~cm$^{-1}$; this mode is not observed due to the
extremely large electronic background.  However, there is a sharp feature at
318~cm$^{-1}$ that is tentatively assigned as the second harmonic of this vibration.
The high-frequency $E_u$ mode, which involves the Co and O atoms, is
calculated to be at $\simeq 628$~cm$^{-1}$ and is observed at 645~cm$^{-1}$.
Although this feature appears relatively insignificant, it possesses significant
oscillator strength (Table~\ref{tab:modes}); it only appears weak because it is
superimposed on a large electronic background.
The two strong features at $\simeq 588$ and 764~cm$^{-1}$ fall into the characteristic
energy range expected for lattice vibrations; however, they do not correspond to
any of the calculated infrared or Raman vibrations.  Both modes, the one at 588~cm$^{-1}$
in particular, narrow considerably and harden at low temperature, ruling out artifacts
from absorptions elsewhere in the optical path as their origin.  These structures are
considerably broader and stronger than expected for infrared-active vibrations.
It appears these features are manifestations of the \emph{c}-axis $A_{2u}$ LO modes,
which have been observed in the cuprates and have the same antiresonant line shape
in the reflectance \cite{reedyk92} (inset of Fig.~\ref{fig:modes}).  For a single
oscillator, $\Omega_0^2 = \epsilon_\infty (\omega_{\rm LO}^2-\omega_{\rm TO}^2)$.
Using the values in Table~\ref{tab:modes} and $\epsilon_\infty \simeq 3.4$ returned
from the fits, $A_{2u}$ LO modes at the correct positions can be obtained using
$\Omega_0 \simeq 700 - 950$~cm$^{-1}$.  This suggests the presence of electron-phonon
coupling where the out-of-plane displacements of the Pd and O atoms allow the in-plane
carriers to couple of the long-range electric field \cite{reedyk92}.

%
% Table II
%
\begin{table}[t]
\caption{The fitted Lorentz oscillator parameters are listed for the four features
observed in $\sigma_1(\omega)$ at 5~K in Fig.~\ref{fig:modes}, and compared with the
calculated frequencies and atomic intensities of PdCoO$_2$ of the infrared-active
modes at the zone center.  All units are in cm$^{-1}$ unless otherwise indicated.}
\begin{ruledtabular}
\begin{tabular}{cc c ccc c cccc}
   & & & \multicolumn{3}{c}{Experiment (5~K)} & & \multicolumn{4}{c}{Theory$^{\rm a}$} \\
   Mode   & (branch) & & $\omega_i$ & $\gamma_i$ & $\Omega_i$ & &
                                             $\omega_{calc}$ & Pd  &  Co  &   O  \\
   \cline{1-2} \cline{4-6} \cline{8-11}
    $E_{u,1}$ & (TO) & & $-$ & $-$  &  $-$ & &  154     & 0.45 & 0.43 & 0.12 \\
    $A_{2u}$  & (TO) & & $-$ & $-$  &  $-$ & &  287     & 0.43 & 0.52 & 0.05 \\
   $2E_{u,1}$ & (TO) & & 318 &  2.3 & 1334 & &  308     &  $-$ & $-$  & $-$  \\
    $A_{2u}$? & (LO) & & 588 &  53  & 6566 & &   $-$    &  $-$ & $-$  & $-$  \\
    $E_{u,2}$ & (TO) & & 645 &  6.7 &  872 & &  628     & 0.00 & 0.28 & 0.72 \\
    $A_{2u}$  & (TO) & & $-$ & $-$  &  $-$ & &  661     & 0.03 & 0.18 & 0.79 \\
    $A_{2u}$? & (LO) & & 764 & 13.5 & 2080 & &   $-$    &  $-$ & $-$  &  $-$ \\
\end{tabular}
\end{ruledtabular}
\footnotetext[1] {This work.}
\label{tab:modes}
\end{table}
%

%
% Summary
%
\section{Conclusions}
The in-plane optical properties of PdCoO$_2$ reveal that the free-carrier
intraband response falls well below the interband transitions, allowing the plasma
frequency to be determined from the \emph{f}-sum rule; the value of $\omega_p \simeq
33\,300$~cm$^{-1}$ is in good agreement with fits to the Drude-Lorentz model, as well as
first-principle calculations.
While the optically-determined scattering rate at room temperature of $1/\tau_{op} \simeq
100$~cm$^{-1}$ is in good agreement with transport measurements, it displays little
temperature dependence, and at low temperature $1/\tau_{op} \gg 1/\tau_{tr}$.
This inconsistency is resolved by assuming Fermi liquid behavior where the scattering rate
varies quadratically with both temperature and frequency; $1/\tau_{op}$ is then the average of
$1/\tau(\omega)\propto \omega^2$ over the region of the free-carrier response, unlike the
$1/\tau(\omega) \propto \omega$ behavior observed in the cuprates.
%
% and indicates PdCoO$_2$ is at best weakly-correlated material.
%
Despite the high conductivity of this material, at least one in-plane infrared-active $E_u$
mode is identified.  The two additional features appear to be manifestations of the
$A_{2u}$ \emph{c}-axis LO modes coupling to the in-plane carriers; the strength and width
of these features suggests that electron-phonon coupling is present. \\

%
% Acknowledgements
%
\begin{acknowledgements}
We would like to acknowledge helpful conversations with Ana Akrap and Jungseek Hwang.
Work at Brookhaven National Laboratory was supported by the Office of Science, U.S. Department
of Energy under Contract No. DE-SC0012704.
\end{acknowledgements}
\vfil

%
%%%%%%%%%%%%%%%%%%%%%%%%%%%%%%%%%%%%%%%%%%%%%%%%%%%%%%%%%%%%%%%%%%%%%%%%%%%%%%
%
% References
%
%\bibliography{references}

\begin{thebibliography}{34}
\bibitem [{MIR()}]{MIR}%
  \BibitemOpen
  \href@noop {} {}\bibinfo {note} {The Mott-Ioffe-Regel limit is the point at
  which the mean free path is equal to a lattice spacing and the Boltzmann
  description of diffusive transport breaks down}\BibitemShut {NoStop}%
\end{thebibliography}
%


%
% Supplementary material
%

\clearpage
\newpage

\newpage
\vspace*{-2.1cm}
\hspace*{-2.5cm}
{
  \centering
  \includegraphics[width=1.2\textwidth,page=1]{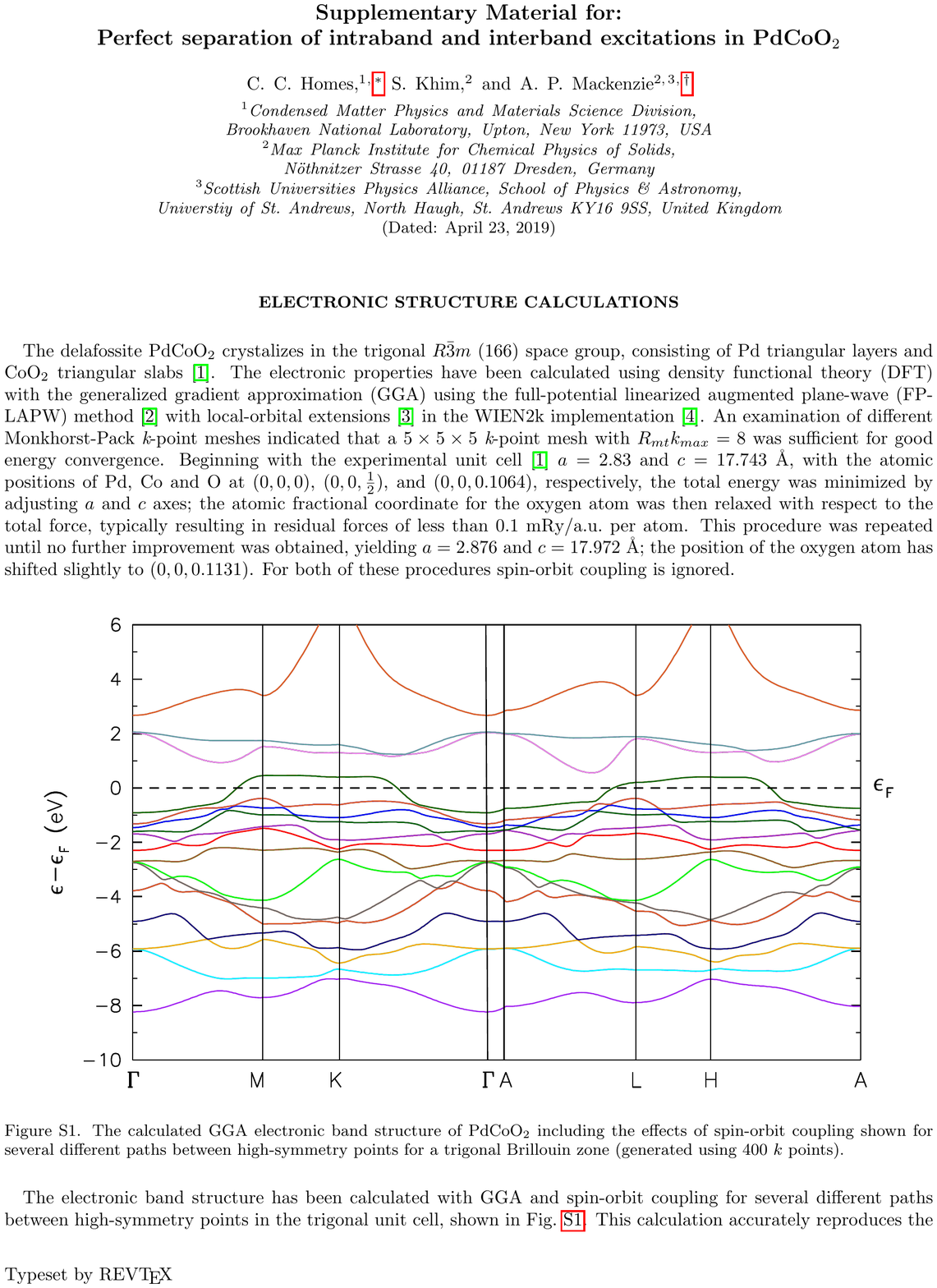} \\
  \ \\
}

\newpage
\vspace*{-2.1cm}
\hspace*{-2.5cm}
{
  \centering
  \includegraphics[width=1.2\textwidth,page=2]{supplemental.pdf} \\
  \ \\
}

\newpage
\vspace*{-2.1cm}
\hspace*{-2.5cm}
{
  \centering
  \includegraphics[width=1.2\textwidth,page=3]{supplemental.pdf} \\
  \ \\
}

\newpage
\vspace*{-2.1cm}
\hspace*{-2.5cm}
{
  \centering
  \includegraphics[width=1.2\textwidth,page=4]{supplemental.pdf} \\
  \ \\
}

\end{document}